%% file: arxiv.tex
\documentclass[onecolumn]{article}
\usepackage[utf8]{inputenc}
\usepackage{amsthm}
\usepackage{amsmath}
\usepackage{url}
\usepackage{graphicx}
\usepackage{subcaption}
\usepackage[margin = 1in]{geometry}
\newcommand{\E}{\mathsf{E}}

\usepackage{mathptmx}
\usepackage{amsfonts}
\newtheorem{theorem}{Theorem}
\newcommand{\eat}[1]{}
\newcommand{\mechanism}{\mathcal{M}}
\usepackage[round]{natbib}
\usepackage{hyperref}
\usepackage[conf={no link to proof}]{proof-at-the-end}

\newtheorem{thm}{Theorem}
\newtheorem*{thm*}{Theorem}

\newtheorem{lemma}[thm]{Lemma}
\newtheorem*{lemma*}{Lemma}

\newcommand{\threshold}{\tau}
\newcommand{\newtau}{v}
\newcommand{\sm}{Appendix}

\begin{document}

\title{Sample and Threshold Differential Privacy: \\ Histograms and applications}
\author{Akash Bharadwaj\thanks{akashb@fb.com} \and Graham Cormode\thanks{gcormode@fb.com}}
\date{}

\maketitle

\input{content}


\clearpage
\renewcommand{\bibname}{References}
\bibliographystyle{apalike}
\bibliography{triehh}

\clearpage

\appendix

\section{Omitted technical material}

\printProofs

\end{document}

%% file: content.tex
\newcommand{\eeps}{e^{-\epsilon}}
\newcommand{\eepspos}{e^{\epsilon}}
\newcommand{\eepsneg}{\eeps}
\allowdisplaybreaks
\begin{abstract}
Federated analytics seeks to compute accurate statistics from data distributed across users' devices while providing a suitable privacy guarantee and being practically feasible to implement and scale. 
In this paper, we show how a strong $(\epsilon, \delta)$-Differential Privacy (DP) guarantee 
can be achieved for the fundamental problem of histogram generation in a federated setting, via a highly practical sampling-based procedure that does not add noise to disclosed data.
Given the ubiquity of sampling in practice, we thus obtain a DP guarantee almost for free, avoid over-estimating histogram counts, and allow easy reasoning about how privacy guarantees may obscure minorities and outliers.
Using such histograms, related problems such as heavy hitters and quantiles can be answered with provable error and privacy guarantees. 
Experimental results show that our sample-and-threshold approach is accurate and scalable.
\end{abstract}

\section{\uppercase{Introduction}}
\allowdisplaybreaks
Building private histograms is a task that underpins a variety of machine learning and data analytics tasks. 
Histograms enable building usable discrete representations, distributions and marginals.
Materializing histograms is also a core subroutine in instantiating graphical models for synthetic data generation~\citep{NIST21}, and hence they support numerous statistical analyses and inference tasks. 
The problem has been heavily studied with an eye for differential privacy, with a number of results shown under various models, such as the central model~\citep{Dwork06,XuZXYY12, DworkR14}, local model~\citep{bassilysmith,wangetal,hadamardresponse} and shuffle model~\citep{ESA20,BalcerCheu20,DUMP20}.

In this paper, we revisit this foundational question, and show how differential privacy can be obtained via a simple sample-and-threshold mechanism, which can be readily implemented in a distributed setting.  
Importantly, all the randomization needed for privacy is derived from the sampling operator: there is no further explicit addition of noise to data disclosures. 
This is particularly beneficial in scenarios when sampling is inherent, i.e., federated settings when only a uniformly chosen fraction of users are contacted. 
In this case, privacy essentially comes ``for free''. 
Equipped with an efficient mechanism for histogram computation, we can apply it to a range of core analytics tasks (quantiles and heavy hitters), which in turn enable a broad spectrum of other computations.  

\paragraph{Our contributions.}
In this paper, we present a histogram mechanism that extends prior work as follows: 
\begin{itemize}
\itemsep0em
\item We show that a simple sample-and-threshold approach provides an $(\epsilon, \delta)$-differential privacy guarantee for histograms. 
    \item We show that the resulting mechanism can also answer heavy hitter, quantile and range queries.
    \item We show that the associated counts provide accurate frequency estimates for items from the input.
    \item Our proofs are compact and self-contained. 
\end{itemize}
In more detail, we show that a Poisson sampling-based approach is sufficient to provide differential privacy. 
The key is to choose a small enough sampling rate to introduce uncertainty, and to prune items with low frequency in the sample, so that the presence of an item in the pruned sample does not indicate exactly how many instances were in the original population. 
While prior work has considered the ability of sampling to amplify the privacy bounds of a differentially private mechanism, in this work we show that sampling itself provides a DP histogram mechanism, generalizing the pioneering work of \cite{TrieHH} on heavy hitters.
Consequently, the sample-and-threshold histogram mechanism can be implemented effectively while requiring very little effort from participating users. 
The chief points of comparison are results in the shuffle model of differential privacy. 
We claim that the sampling step is arguably simpler than many shuffle approaches (which require users to perturb their inputs, or add additional ``chaff'' messages to mask their values), while being of equivalent complexity to implement the server-side aggregation of messages. 
Deployed federated systems~\citep{BonawitzEGHIIKK19,Papaya} already implicitly sample from a large collection of eligible users, so the mechanism does not introduce any significant additional overhead or error. 

\section{\uppercase{Preliminaries}}

We consider the case where there are $n$ individuals who each hold 
a value $x_i$, so that the collection of all user inputs defines a dataset $D$. 
Our goal is to construct a histogram of the frequency distribution according to a
fixed set of buckets $B$. 
For convenience, we assume that each input $x_i$ is already mapped into its corresponding bucket, and that the buckets are indexed by integers, so that each $x_i \in [B]$. 
We will describe a randomized mechanism, $\mechanism$, that can process datasets $D$ to give a distribution over output histograms, $H$. 

The objective is to ensure that a sampled output histogram, $H$,  is close to the true histogram $H^*$, while ensuring that the output meets $(\epsilon, \delta)$-differential privacy (DP) \citep{DworkR14}. 
Formally, we require
\begin{equation}
\Pr[ \mechanism(D) \in \mathbb{H}] \leq \exp(\epsilon)\Pr[\mechanism(D') \in \mathbb{H}] + \delta
\label{eq:dp}
\end{equation}
\noindent
for any subset of possible output histograms $\mathbb{H}$ and
for neighboring inputs $D$, $D'$ that differ in the input value of one individual. 
As usual for $(\epsilon, \delta)$-DP, we expect $\delta$ to be small, typically much less than $1/n$. 

\paragraph{Computational model.}
Our mechanism is designed to operate in a federated (distributed) setting, where each client sends a message based on their input to a server, which then combines this information before reporting it to an analyst. 
Specifically, the server aggregates the messages to produce the multiset of values reported (i.e., builds the frequency histogram of messages), and deletes some values which fall below a threshold $\threshold$. 
This model sits between the shuffle and centralized DP model: 
the  procedure is conceptually similar to the `shuffling' procedure, but with the minor additional step of removing small counts; meanwhile, it is easy to implement in the central DP model with a trusted aggregator~\citep{ESA20}. 
Many shuffling protocols are based on clients following a locally differentially private (LDP) protocol, based on a high setting of the privacy parameter $\epsilon$; the shuffling then `amplifies' the privacy to give a tighter DP guarantee on the output. 
Compared to many shuffle and LDP protocols, our approach is very compact: it only sends information on the items held by the client, rather than the size of the domain from which those items are drawn.

To fully achieve the benefits of the sample-and-threshold model, 
we assume for convenience that there is an entity which aggregates the data, similar to a shuffler in the shuffle model. 
For a shuffler, applying the threshold would be a trivial final step before the shuffler releases the histogram. 
Indeed, we anticipate that this would be natural to do in any system that implements aggregation via secure hardware (e.g., Intel SGX extensions~\citep{IntelSGX}). 
Then the data analyst only sees output under differential privacy, and is shielded from seeing any intermediate results without a formal privacy property.  
Alternatively, the aggregation and thresholding step could be performed using techniques from multiparty computation on shares of the input gathered by two or more servers, wherein secure comparison to a public constant is a relatively lightweight operation~\citep{Nishide:Ohta:07,VeugenBHE15}. 
The model can also be compared to the early notion of ``$k$-anonymity'', where the output is constrained so that every output item corresponds to at least $k$ individuals in the input~\citep{Samarati:Sweeney:98}.  
Here, we obtain $k$-anonymity for $k = \threshold$, the threshold value. 
Although $k$-anonymity has been criticized as a weak privacy notion, it carries an intuitive appeal for many lay users, and here we show that in this case we also achieve differential privacy. 

\section{\uppercase{Sampling-Based Histograms}}

In the ($B$-bucket) histogram problem, each client $i$ holds a single item $x_i$ corresponding to a bucket $b_i \in [B]$, and our aim is to produce a private histogram of item frequencies, 
such that a frequency associated with $x$ in the private histogram approximates the frequency of $x$ over the input distribution. 

\subsection{Main Sampling Results}

The algorithm is based on Bernoulli sampling. 
Each client out of $n$ is sampled with probability $p_s = m/n$, so the expected size of the sample is $m$ (we later discuss different ways to implement this sampling). 
Our subsequent analysis will relate the sampling rate to the privacy parameter $\epsilon$ in order to give a required privacy guarantee. 
The algorithm makes use of a threshold $\threshold$, so that items whose sampled counts are at least $\threshold$ are reported in the histogram, while items 
whose count falls below $\threshold$ are omitted from the histogram. 
Note then that the mechanism introduces no spurious items into the output: any item which is not present in the input cannot appear in the output histogram. 
Hence, the size of the message from each client need be no bigger than its input size. 
In addition, the error bounds of the algorithm are independent of the dimensionality of the underlying histogram, $B$.

We next give a bound on the ratio of probabilities of seeing the same output on neighboring inputs, which allows us to state our main result. 

\begin{theoremEnd}[restate]{lemma}
Given two neighboring inputs $D$, $D'$, such that $D$ differs in one item from $D'$, 
the ratio of probabilities of seeing a cell with a given count $\newtau$ is bounded by $\frac{(1 - p_s)k}{k - \newtau}$, 
where $k$ is the number of copies of the given item in input $D$
and $k > \newtau$. 
\label{lem:rationew}
\end{theoremEnd}

\begin{proofEnd}
The case to focus on is when input $D$ has one extra copy of a particular item compared to $D'$, at some intermediate stage of the algorithm. 
%
We can directly compute the probability of picking $\newtau$ out of the $k$ occurrences of the item by the Binomial formula as
${ k \choose \newtau} p_s^\newtau (1 - p_s)^{k - \newtau}$.
Our goal is to bound the ratio of probabilities of seeing a count of $\newtau$ copies of the item in the output of $D$, who has $k$ copies of the item, and of $D'$ who holds $k-1$ copies.   
Then this ratio of probabilities is given by
\begin{align*}
    \frac{ { k \choose \newtau } p_s^\newtau (1 - p_s)^{k - \newtau} }{ { k-1 \choose \newtau } p_s^\newtau (1 - p_s)^{k - \newtau - 1} }
    & = 
    \frac{ k!  \newtau! (k - 1 - \newtau)! (1 - p_s) }{ (k-1)! \newtau! (k - \newtau)!}
   \\& = \frac{(1-p_s)k}{k - \newtau}
   \end{align*}\end{proofEnd}

The proof of this claim, and of some other technical lemmas, is deferred to \sm. 

\begin{theorem}
Sample-and-threshold achieves $(\epsilon, \delta)$-differential privacy provided 
$\epsilon$ and $\delta$ satisfy
$\eeps \leq 1 - p_s$ and
$\delta \leq \exp(-\frac{\tau}{q} D(q\|p))$
for $q = 1 - \eeps(1 - p_s)$, where
$D(q\|p)$ denotes the KL divergence. 
\label{thm:main}
\end{theorem}

\begin{proof}
We must show that the ratio $\frac{(1-p_s)k}{k-\newtau}$ is between $\eepsneg$ and $\eepspos$ except with some small probability. 
For the lower bound, we have
\begin{equation}
    \eeps \leq \frac{(1-p_s)k}{k-\newtau} 
    \label{eq:lb}
\end{equation}
for any $\newtau \geq 0$, which is satisfied if we ensure $p_s \leq 1 - \eeps < 1$ (since
$v=0$ is the worst case).
Rearranging the upper bound, we obtain
\begin{equation}
    \newtau \le k\eeps( \eepspos - 1 + p_s) = k (1 - \eeps + \eeps p_s)
    \label{eq:newtau}
\end{equation} 

First, note that since $p_s < 1$, we have $p_s (1 - \eeps) < 1 - \eeps$, and 
so $p_s < 1 - \eeps + \eeps p_s$. 
Hence, the bound in \eqref{eq:newtau} is greater than $k p_s$, the mean value. 
In fact, 
 $(1 - \eeps + \eeps p_s) < 1$ since $p_s < 1$, and so we can define the probability
\begin{equation}q = ( 1 - \eeps + \eeps p_s).\label{eq:q}\end{equation}
We will thus obtain $(\epsilon, \delta)$-differential privacy provided that we can bound the probability of choosing
a $\newtau$ that is more than $kq$. 
Recall that applying the threshold means that if the sampled number of items is below $\tau$ then it is rounded down to 0. 
So the ``bad event'' is when more than $\max(kq, \tau)$ out of $k$ copies are sampled. 
The number of sampled copies is a Binomial distribution with $k$ trials and success probability $p_s$, $B(k, p_s)$. 
Hence, we analyze $\Pr[ B(k , p_s) >\max(kq, \tau)]$.

We perform a case split, based on the value of $k$. 

\textbf{Case 1}.  $k \leq \frac{\tau}{q}$. 
In this regime, we have that $\newtau = \max(k q, \tau) = \tau$, and so we seek 
$\Pr[ B(k, p_s) > \tau]$. 
Clearly, this is monotonic in $k$: increasing $k$ only makes it easier to achieve more than $\tau$ successes. 
Hence, it suffices to consider the case 
$k = \frac{\tau}{q}$ and $\newtau = \tau$. 

\textbf{Case 2}. $k \geq \frac{\tau}{q}$
We apply the Chernoff-Hoeffding bound, which states that 
\begin{align}
    \Pr[ B(k, p) > kq ] & = \Pr[ B(k, 1-p) < k - kq]  \nonumber \\
    & \leq \exp(-k D( q \| p ) )
\label{eq:ch}
\end{align}
where $D(q\|p)$ denotes the KL divergence (relative entropy) between the (Bernoulli) distributions with parameters $q$ and $p$. 
We can observe that \eqref{eq:ch} is decreasing as a function of $k$ (since $D(q\|p)$ is non-negative for all $p, q$), and
so the tightest bound is for the smallest value of $k = \frac{\tau}{q}$. 

Combining the observations of Case 1 and Case 2, we apply \eqref{eq:ch} for $k = \frac{\tau}{q}$ to get
$\delta \leq \exp(-\frac{\tau}{q} D(q\|p))$.
\end{proof}

The best bound on $\delta$ will be obtained by using the expression from this Theorem (or the exact Binomial distribution) at 
$k = \frac{\tau}{q}$.
However, to give a simpler expression, we will provide a slightly looser bound, as follows. 

\begin{theoremEnd}[restate]{lemma}
If we set the sampling rate $p_s = \alpha(1 - \eeps)$ for some $0 < \alpha \leq 1$ and $\epsilon \leq 1$,
then sample-and-threshold achieves $(\epsilon, \delta)$ differential privacy 
for $\delta = \exp(-C_\alpha \tau)$, 
where $C_\alpha = \ln(1/\alpha) - 1/(1+\alpha)$.
\end{theoremEnd}

\begin{proofEnd}
From the definition of $q$ in \eqref{eq:q}, we have
$ 1- q = \eeps - \eeps p_s = (1 - p_s)\eeps $.
Then
\begin{align*}
    D(q\|p_s) & = q \ln \frac{q}{p_s} + (1-q) \ln \frac{1-q}{1-p_s} \\
 & =    
   q\ln\left( \frac{1 - \eeps(1-ps)}{p_s}\right) 
    +  
   (1-q) \ln \eeps \\
 & = 
\textstyle
   q\ln\left( \frac{1}{p_s}(1  - \eeps + \eeps p_s) \right) 
   -  
   \epsilon(1-q)
\end{align*}
To simplify this, we use $p_s = \alpha(1 - \exp(-\epsilon))$.
Then 
\begin{align*}
    D(q \|p_s) & = q \ln \left( \frac{(1 + \alpha\eeps)(1 - \eeps)}{\alpha(1 - \eeps)}\right) 
     - \epsilon(1-q) \\
     & = \textstyle q \ln \frac{1 + \alpha \eeps}{\alpha} 
    - \epsilon(1-q)
\end{align*}
Using the Hoeffding-Chernoff bound, we get 
\begin{align}
\nonumber
    \delta & \textstyle
 \leq \exp( - \frac{\tau}{q} D(q\|p)) = \exp(-\tau (\ln \frac{1 + \alpha \eeps}{\alpha} 
    - \frac{1-q}{q}\epsilon))  \\
    & \textstyle \leq \exp(-\tau (\ln( 1/\alpha) -  \frac{1-q}{q}\epsilon))
    \label{eq:increasing}
\end{align}

We can observe that \eqref{eq:increasing} is decreasing as a function of $\epsilon$, by expanding out the definition of 
\begin{align}
    \frac{1-q}{q} \epsilon  \nonumber
    & = 
    \frac{\eeps(1 - \alpha + \alpha \eeps)}{(1 - \eeps)(1 + \alpha \eeps)} \epsilon
    = \frac{\epsilon}{\eepspos - 1} \frac{1 - \alpha + \alpha \eeps}{1 + \alpha \eeps}  \\
    & = \frac{\epsilon}{\eepspos - 1} \left( 1 - \frac{\alpha}{ 1 + \alpha \eeps}\right) 
\label{eq:decreasing}
\end{align}

Observe that 
$\frac{-1}{1 + \alpha \eeps}$ is decreasing as a function of $\epsilon$, and so the term
$( 1  - \frac{\alpha}{1 + \alpha\eeps})$ is also decreasing. 
To see the behavior of $\frac{\epsilon}{\eepspos - 1}$, we differentiate wrt $\epsilon$:
\[
\frac{d}{d\epsilon} \frac{\epsilon}{\eepspos - 1} = -\frac{\eepspos(\epsilon - 1) + 1}{(\eepspos-1)^2}
\]
This derivative is clearly negative for $\epsilon \ge 1$, indicating a decreasing behavior. 
For $0 < \epsilon < 1$, we study the numerator 
$f(\epsilon) = -(\eepspos(\epsilon - 1) + 1)$.
Then $f(0) = 0$ and $f(1) = -1$. 
In between, we have 
\[
\frac{d}{d\epsilon} f(\epsilon) = - \frac{d}{d\epsilon} \eepspos(\epsilon - 1) + 1 = -\epsilon \eepspos
\]
which is negative for $\epsilon > 0$, allowing us to conclude that 
$\frac{\epsilon}{\eepspos -1}$ is decreasing for all $\epsilon > 0$. 

Putting this all together, we conclude that 
$\exp(-\tau(\ln(1/\alpha) - \frac{1-q}{q}\epsilon)$ is decreasing as a function of $\epsilon$, and so achieves its greatest value in the limit as $\epsilon$ approaches 0. 
Considering~\eqref{eq:decreasing}, 
we have
\begin{align*}
\lim_{\epsilon \rightarrow 0} \frac{1-q}{q}\epsilon & = 
    \lim_{\epsilon \rightarrow 0} \frac{\epsilon}{\eepspos - 1} \left( 1 - \frac{\alpha}{ 1 + \alpha \eeps}\right) \\
    & = \lim_{\epsilon \rightarrow 0} \frac{\epsilon}{\eepspos - 1} 
    \left( \lim_{\epsilon \rightarrow 0}  1 - \frac{\alpha}{1 + \alpha \eeps}\right)\\
    & = 1 \cdot \left(1 - \frac{\alpha}{1+\alpha}\right) 
    = \frac{1}{1+\alpha}
\end{align*}
Substituting this into \eqref{eq:increasing}, we find
\[ 
\textstyle
\delta \leq \exp\left(-\tau\left(\ln(1/\alpha) - \frac{1}{1+\alpha}\right)\right) = \exp(-\tau C_\alpha) 
\]
where we define $C_\alpha = \ln(\frac{1}{\alpha}) - \frac{1}{1+\alpha} = O(\ln(\frac{1}{\alpha}))$.
\end{proofEnd}

\eat{
To bound the expression for $\delta$, first we observe that (for $\epsilon \leq 1$):
\begin{align*}
    \frac{\tau}{q} & = \frac{\tau}{1 - \eeps + \alpha(1 - \eeps) \eeps} \\
    & = \frac{\tau}{(1 - \eeps) (1 + \alpha \eeps)} \\
    & = \frac{\tau}{\Theta(\epsilon(1 + \alpha))} = \Theta\left(\frac{1-\alpha}{\epsilon}\right) \tau . 
\end{align*}
To bound $D(q\|p)$, we observe that
\begin{align*}
\epsilon(1 - p_s)\eeps & \leq \epsilon \eeps  \qquad\text{and}  \\
\textstyle
     (1 - \eeps)(1 + \alpha \eeps) \ln \frac{1 + \alpha \eeps}{\alpha}  
  &  = \textstyle \Theta(\epsilon (1 + \alpha \eeps) \ln \frac{1}{\alpha})
\end{align*}
Combining these bounds, we obtain
\begin{align*}
    \delta & \leq \exp\left(-\frac{\tau}{q} D(q\|p_s)\right) \\
    & = \exp\left(-\Theta\left(\frac{\tau(1-\alpha)}{\epsilon} \epsilon ( (1 + \alpha \eeps) \ln 1/\alpha - \eeps) \right)\right) \\
    & = \exp\left(-\Theta\left( \tau(1 - \alpha) ((1 + \alpha \eeps) \ln 1/\alpha - \eeps) \right) \right)
\end{align*}
As $\epsilon$ decreases towards 0, we can approximate the terms $\exp(-\epsilon)$ by 1, and obtain for $\alpha < 1$ that
\begin{align*} 
  \delta & 
    < \exp\left(-\tau \cdot \Theta\left( (1 - \alpha) ( (1 + \alpha) \ln(1/\alpha) + 1) \right) \right)  \\
 & = \exp(-\tau \cdot \Theta(\ln(1/\alpha))) = \alpha^{\Theta(\tau)}
\end{align*}
}

Hence, for $\alpha$ small enough, we have
$\delta \leq \exp(-\tau O(\ln(1/\alpha))) = \alpha^{O(\tau)}$, 
decaying exponentially as a function of $\tau$. 

This analysis provides guidance on how to set the sample-and-threshold parameters in order to obtain a desired $(\epsilon, \delta)$ guarantee: 
we first set the sampling rate $p_s = \alpha(1 - \eeps) = \theta(\alpha \epsilon)$, and
the threshold $\tau = \frac{1}{C_\alpha} \ln(1/\delta)$.
Equivalently, given $p_s$ and $\alpha$, we obtain
$\epsilon = \ln(\frac{\alpha}{\alpha - p_s})$.
Typically, we will set $\alpha$ to be a constant (say, $1/5$), which will ensure 
$(\epsilon, \delta)$ privacy when sampling with probability $\epsilon/5$, and choosing an appropriate threshold to obtain a small enough $\delta$. 

Concretely, for $\epsilon = 1$ and $\alpha = 1/6$, the sampling rate is $p_s = 0.105 \approx 0.1$ and, choosing $\tau = 20$,  $\delta < 10^{-8}$ using $C_\alpha = 0.935$.

\subsection{Fixed sized sampling}

For practical efficiency, we would often like to work with a fixed size sample. 
However, the above histogram protocol performs Poisson sampling instead.
The reason is that if the fixed size of the sample, $m$, is known, then we are effectively also releasing the number of samples that were suppressed by the $\threshold$ threshold (by adding up the released counts, and subtracting from $m$). 
This potentially leaks information. 
Consider the case where $D'$ contains $n$ copies of the same item, while $D$ contains $n-1$ copies of the same item, and one unique item. 
With probability $m/n$, the mechanism on input $D$ samples the unique item along with $m-1$ other items, and so produces an output of size $m-1$. 
But on input $D'$, there is zero probability of producing an output smaller than $m$. 
This forces $\delta \geq m/n$, which is typically too large for $(\epsilon, \delta)$-DP (we usually seek $\delta \ll 1/n$). 

Performing Poisson sampling with $p_s = m/n$ addresses this problem: the expected sample size is the same, but we no longer leak the true size of the sample before thresholding.  
Indeed, we can see that the (observable) size of the sample is differentially private: given two inputs $D$ and $D'$ such that $D$ has one additional unique item, the distribution of sample sizes are close, applying Theorem~\ref{thm:main} to the samples.

Implementing Poisson sampling may appear costly: naively, the server would contact $n$ clients instead of $m$, where we expect $n \gg m$. 
However, we can perform the sampling by contacting much fewer clients, since the size of the sample is tightly concentrated around its expectation.  

\begin{theoremEnd}{lemma}
Sampling $m + O(\sqrt{m})$ clients is sufficient to apply the sample-and-threshold mechanism, 
with high probability. 
\end{theoremEnd}

\begin{proof}
Observe that, with high probability, the size of the (Poisson) sample will be close to expected value of $m$. 
In particular, by a Chernoff bound, the probability that the sample size is more than $c\sqrt{m}$ larger than $m$ is
\[ \Pr[ s > (1 + c/\sqrt{m})m] \leq \exp(-c^2/3)  . \]
Hence, for $c$ a suitable constant (say, 10), this probability is negligibly small.  
To realize this sampling, we contact a fixed size number of clients $s = m + c\sqrt{m}$, and then have each client perform a Bernoulli test on whether to participate: with probability $m/s$, it participates, otherwise it abstains. 
An abstaining client can, for example, vote for a unique element (e.g., an item based on a hash of its identifier), and so be automatically discounted from the protocol, without revealing this information to the aggregator. 
\end{proof}

\subsection{Accuracy Bounds}

The histogram produced by the mechanism is ultimately based on sampling and pruning.
For any item whose frequency is sufficiently above the pruning threshold, 
its frequency within the histogram is an (almost) unbiased estimate of its true frequency. 
There is a small gap, since even for an item with high frequency, there is a small chance that it is not sampled often enough, and so its estimate will fall below the threshold $\threshold$ (in which case we do not report the item). 

\paragraph{Probability of omitting a heavy item.}
We first consider the probability that a frequent item is not reported by the algorithm. 
\begin{theoremEnd}[restate]{lemma}
The sample and threshold histogram protocol omits an item whose true absolute count is $W$ with probability
at most $\exp(-(\frac{Wm}{n} - \threshold)^2\frac{n}{2Wm})$. 
\end{theoremEnd}

\begin{proofEnd}
For an item with (absolute) frequency $W$ out of the $n$ input items, it is reported 
if the number of sampled occurrences exceeds $\threshold$. 
We can apply a multiplicative Chernoff-Hoeffding bound to the random variable $X$ that counts the number of occurrences of the item. 
Now, the probability of each sample picking the item is $W/n$, and the expected number in the sample is
$Wm/n > \threshold$. 
For convenience, we will write $w = Wm/n$ for this expectation. 
We have that\footnote{Here we are sampling without replacement.  However, bounds for sampling with replacement are still valid here.}%
\begin{align*}
    \Pr[ X \leq \threshold] & = \Pr\left[ X \leq \frac{\threshold}{w}w\right] \\ & = 
    \Pr\left[X  \leq \left(1 - \frac{w - \threshold}{w}\right)\E[X] \right]  \\ & = \exp\left(-\frac{(w-\threshold)^2}{2w}\right)
\end{align*}\end{proofEnd}
\paragraph{Numeric Example.}
When $w := Wm/n$ is sufficiently bigger than $\threshold$, this gives a very strong probability.  
For example, consider the case $n=10^6$, $\epsilon = 1$, and we set $p_s = 0.1$ and $\threshold = 20$ to obtain a $\delta$ of $10^{-8}$. 
The expected sample size $m = 10^5$, and for an item that occurs 0.1\% of the time in the input, we expect to sample it
$w = 100$ times. 
This gives a bound of $\exp(-32) < 10^{-13}$ that such an item is not detected.

\paragraph{Frequency estimation bounds.}
More generally, we can use the (relative) frequency of any item in the histogram as an estimate for its true occurrence  in the population. 

\begin{theoremEnd}{lemma}
We can estimate the (relative) frequency of any item whose relative frequency is $\phi$ within $\gamma$ relative error with probability 
$O(\exp(-\gamma^2 \phi m))$.
\end{theoremEnd}

\begin{proof}
Applying a multiplicative Chernoff-Hoeffding bound, we have for $\gamma < 1$, 
\[ 
\Pr[ |X - \mu| > \gamma\mu] = 2\exp( - \gamma^2 \mu/3) = \beta \]
Rearranging, we obtain 
$\mu = \frac{3}{\gamma^2} \ln (1/2\beta)$. 
Suppose we aim to find all items whose frequency is at least $\phi$, and estimate their frequency with relative error at most $\gamma$. 
Then we have $\mu = \phi m = \frac{3}{\gamma^2} \ln (1/2\beta)$. 
\end{proof}

\paragraph{Numeric Example.}
We can substitute values into this expression to explore the space. 
For example, if we set
 $p_s = 0.1$, $\ln (1/2\beta) = 10$, $\threshold = 10$ and $\gamma = 1/\sqrt{10}$, then 
we obtain $\phi = 3 \times 10^3/n$ --- in other words, provided $n > 3 \times 10^5$, we can accurately find 
estimates of frequencies that occur 1\% of the time (except with vanishingly small probability). 

\paragraph{Remark.}
It is instructive to compare these bounds to those that hold for the shuffle model. 
According to \cite{BalcerCheu20}, 
addition of appropriately parameterized Bernoulli random noise to reports from $n$ clients yields
$(\epsilon, \delta)$-DP, with error that scales as $O(\frac{1}{\epsilon^2 n} \log (1/\delta))$ for $\epsilon \leq 1$, provided $n$ is large enough. 
Expressing our bound on the estimate of any frequency, we obtain error $O(1/\sqrt{m}) = O(\sqrt{\frac{1}{\epsilon n}})$ from sampling, plus
error from rounding small values down to zero, which is bounded by $O(\threshold/m) = O(\ln(1/\delta)/m) = O(\ln(1/\delta)/(\epsilon n))$. 
Naively, it might seem that the shuffle bounds are preferable, due to the stronger dependence on $n$ ($O(1/n)$ vs.\ $O(1/\sqrt{n})$). 
However, this misses the point that in practical federated computing settings, the server can contact only a fixed size cohort of $m$ clients out of a much larger (and sometimes unknown) population $n$.
For example, Google's GBoard is trained with batches of 200 clients at a time~\citep{Gboard}; Meta's FedBuff trains with tens to thousands of clients per round~\citep{FedBuff}; while $m$ is set to $O(\sqrt{n})$ for heavy hitter discovery by~\cite{TrieHH}.
In such cases, results in both the shuffle and sample-and-threshold paradigms incur \textit{the same} sampling error of 
$O(1/\sqrt{m})$. 
Then shuffling introduces additional noise of $O(\frac{1}{\epsilon^2 m} \log (1/\delta))$, whereas sample-and-threshold
incurs zero additional noise on items that exceed the $\threshold$ threshold, and at most
$O(\ln(1/\delta)/m)$ on small items. 
Hence, we argue that when shuffling implicitly samples from the input, the sample-and-threshold approach has superior error guarantees. 
We confirm this observation empirically in Section~\ref{sec:expts}, where we compare accuracy of both approaches while sampling the same expected number of clients. 


\section{\uppercase{Heavy hitters and Quantiles via Histograms}}

\subsection{Heavy Hitters}
We next show how to use the basic histogram protocol to find the (hierarchical) heavy hitters from the input. 
This result follows the outline and notation of the TrieHH algorithm \citep{TrieHH}, to allow easy comparison. 

The heavy hitters algorithm proceeds over $L$ levels, to build up a trie of depth $L$. 
At each level, we materialize a histogram of those prefixes of items from the input that extend the current trie. 
This allows us to add items to the current trie based on the threshold $\threshold$, and include the observed count of each prefix for each node in the trie, provided it is more than $\threshold$. 
We can view the TrieHH protocol as materializing a histogram at each level, with progressively finer cells. 
In the protocol as originally described, cells whose ancestor in a previous level did not exceed the $\threshold$ threshold are not eligible for consideration. 
However, the privacy proof still applies if we do not enforce such restrictions. 
We denote our version of the protocol using the new histogram protocol as TrieHH++, to indicate that the trie is augmented with count information. 

\begin{theoremEnd}[restate]{lemma}
\label{lem:hh}
The TrieHH++ protocol using $L$ sample-and-threshold 
histograms with $(\epsilon, \delta)$-DP
achieves an overall guarantee of 
$(L\epsilon, L\delta)$-DP. 
\end{theoremEnd}

The essence of the proof is that the output of the algorithm is the $L$-fold composition of a differentially private mechanism, with some post-processing.  
By the differential privacy of the basic histogram protocol (Theorem~\ref{thm:main}), the result follows. 

\begin{proofEnd}
In more detail, 
we can view the protocol as publishing a histogram at each level, where the granularity of the cells is refined in each round. 
The protocol enforces that if a prefix is not included at a particular level, then none of its extensions are published in any subsequent level. 
However, we can view this as ``post-processing'', and analyze the simpler algorithm that does not enforce this constraint. 
Applying Theorem~\ref{thm:main}, we have that each round satisfies $(\epsilon_i, \delta_i)$-DP for some $\epsilon_i$ and $\delta_i$.

Then we argue that the output of the full protocol is the $L$-fold composition of the mechanisms $M_i$. 
Assuming $\epsilon_i = \epsilon'$ and $\delta_i = \delta'$ for all $i$, then 
using basic composition, we obtain a bound of $(L\epsilon', L\delta')$-differential privacy, leading to the result stated in the theorem claim. 
For $\epsilon_i = \epsilon' < 1$, we can also obtain a tighter bound, of 
$(L\epsilon'^2 + \epsilon' \sqrt{L \log 1/(\delta' L)}, 2L \delta')$ using advanced composition~\citep{DworkR14}. 
\end{proofEnd}

\paragraph{Remark.}
If the objective is only to find the heavy hitters, then the factor of $L$ can be dropped from these bounds. 
That is, instead of proceeding in rounds, we simply apply the basic histogram protocol to the full inputs, and report the items which survive the thresholding process (along with their associated counts if desired). 
Following the above analysis, the resulting output is $(\epsilon, \delta)$-differentially private.
The motivation for having $L$ rounds given by~\cite{TrieHH} is to reduce the exposure of the server to private information: it only observes prefixes from clients that extend shorter prefixes that are already known to be popular.  However, this does not impact on the formal differential privacy properties of the output. 



\eat{
\begin{figure*}[t]
\subcaptionbox{Accuracy: $\epsilon = 0.1, B = 100$ \label{fig:bin:eps0.1b100}}{
\includegraphics[width=0.25\textwidth]{figs/Absolute error at eps=0.1, domain_size=100, population=1e+06, sample_size=1806, distribution=Binomial, top_k=50.png}}%
\subcaptionbox{Accuracy: $\epsilon = 1.0, B = 100$ \label{fig:bin:eps1.0b100}}{
\includegraphics[width=0.25\textwidth]{figs/Absolute error at eps=1.0, domain_size=100, population=1e+06, sample_size=29509, distribution=Binomial, top_k=50.png}}%
\subcaptionbox{Accuracy: $\epsilon = 0.1, B = 10000$ \label{fig:bin:eps0.1b10000}}{
\includegraphics[width=0.25\textwidth]{figs/Absolute error at eps=0.1, domain_size=10000, population=1e+06, sample_size=1806, distribution=Binomial, top_k=5000.png}}%
\subcaptionbox{Accuracy: $\epsilon = 1.0, B = 10000$ \label{fig:bin:eps1.0b10000}}{
\includegraphics[width=0.25\textwidth]{figs/Absolute error at eps=1.0, domain_size=10000, population=1e+06, sample_size=29509, distribution=Binomial, top_k=5000.png}}
\caption{Top-$k$ accuracy results for Binomially distributed data}
\label{fig:bin}
\end{figure*}
}

\subsection{Quantiles}

Finding the quantiles is a common analytics task to describe the distribution of values held by the clients.
We describe two approaches to finding quantiles, both making use of our histogram mechanism. 

\paragraph{Single quantiles via interactive search.}
Given client inputs which fall in the range $[0, 1]$, we seek a value $f$ such that 
the fraction of clients whose value is below $f$ is (approximately) $\phi$. 

\begin{theoremEnd}[restate]{lemma}
Given a $\phi > \threshold/m$, 
we can use $h$ applications of the $(\epsilon, \delta)$-DP histogram mechanism to 
find a value $f$ such that 
$f \pm 2^{-h}$ is a $\phi \pm O(m^{-1/2})$ quantile, 
with $(h\epsilon, h\delta)$-DP. 
\end{theoremEnd}

\begin{proofEnd}
The quantile query can be carried out by a binary search: 
we begin by creating a histogram with buckets $[0, \frac12], [\frac12, 1]$, and recursively try different split points $[0, t], [t, 1]$ until 
we obtain a result with approximately a $\phi$ fraction of points in the first bucket, at which point we can report $t$ as the $\phi$-quantile. 
Provided $\phi$ is sufficiently larger than $\threshold/m$ (and smaller than $1 - \threshold/m$), then we are unlikely to hit any cases where a bucket count is removed. 
As a result, the error will primarily the error from sampling, which is $O(1/\sqrt{m})$~\citep{Lane:03}, plus the error from rounding, which is $2^{-h}$ if we perform $h$ steps of binary search.  
That is, we find a result $t$ such that there is a point 
in the range $[t-2^{-h}, t+2^{-h}] := t \pm 2^{-h}$ that dominates $\phi \pm O(1/\sqrt{m})$. 
The privacy guarantee is $(h \epsilon, h\delta)$, from the composition of $h$ $(\epsilon,\delta)$-DP histograms.  
\end{proofEnd}

This approach is very effective for single queries, but is less desirable when we have a large number of quantile queries to answer in parallel, in which case the hierarchical histogram approach is preferred. 

\paragraph{Quantiles and range queries via hierarchical histograms.}
A common technique to answer quantile and range queries in one-dimension is to make use of hierachical histograms: histograms with geometrically decreasing bucket sizes, so that any range can be expressed as the union of a small number of buckets.  
We can observe that the trie built as part of the TrieHH++ protocol is exactly such a hierarchical histogram, and hence can be used to answer quantile queries, with the same privacy (and similar accuracy) guarantees as for heavy hitters. 

Assume again that each client has an input value in the range $[0,1]$ (say).
We can interpret these as prefixes, corresponding to subranges. 
If the branching factor of the trie, $\beta$, is set to 4, then the value
$\frac{1}{3}$ falls in the range $[0.25, 0.5]$ for a prefix of length 1; 
and in the range $[\frac{5}{16}, \frac{6}{16}]$ for a prefix of length 2. 
Using this mapping of values to prefixes
the algorithm outputs the (DP) trie with weights on nodes as before. 

To answer a range query $[0, r]$, we decompose the range greedily into chunks that can be answered by the trie. 
For example, if $\beta=4$, and we want the range $[0,0.7]$, 
we find the chunks
$[0, \frac14], [\frac14, \frac24]$ at level 1;
$[\frac{8}{16}, \frac{9}{16}], [\frac{9}{16}, \frac{10}{16}], [\frac{10}{16}, \frac{11}{16}]$ at level 2; 
and so on. 
If the trie has $L$ levels, then any prefix query can be answered with $L(\beta - 1)$ probes to the histograms ($\beta - 1$ for each level). 
Moreover, quantile queries are answered by finding range queries whose weight is (approximately) the desired quantile $\phi$. 

Due to the pruning, we will not have information on any ranges whose sampled weight is less than $\threshold$, corresponding to 
a $\threshold/m$ fraction of mass.  
This will give a worst-case error bound of $(\beta -1)\threshold/m$ per level, and so $L(\beta-1)\threshold/m$ over all levels. 
Based on our setting of $m$ proportional to $\epsilon n / (L\threshold)$, we obtain a total error of 
$ (\beta - 1)(L\threshold)^2/\epsilon n$.
In summary, as a consequence of the privacy guarantee from Lemma~\ref{lem:hh}, we can state: 


\begin{lemma}
We can build a set of $L$  $(\epsilon, \delta)$-DP histograms to answer any quantile query $\phi$ 
to find a value $f$ satisfying $(L\epsilon, L\delta)$-DP such that 
$f \pm 2^{-L}$
is a $\phi \pm O((L\threshold)^2/\epsilon n)$ quantile. 
\end{lemma}

\paragraph{Numeric Example.}
Picking similar test values as above shows that this can give reasonable accuracy for $n$ large enough. 
For $\threshold = 10$, $L = 10$, $\beta = 2$, $\epsilon = 1$, the error bound yields
$10^4/n$.  
So for $n > 10^6$, we obtain rank queries (and quantiles) in this space with error around $0.01$.

\begin{figure*}[t]
\centering
\includegraphics[trim=580 25 580 25,clip,width=0.9\textwidth]{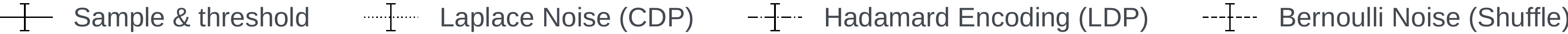}
\subcaptionbox{ Binomial, $B=2^6$ }{\includegraphics[width=0.33\textwidth]{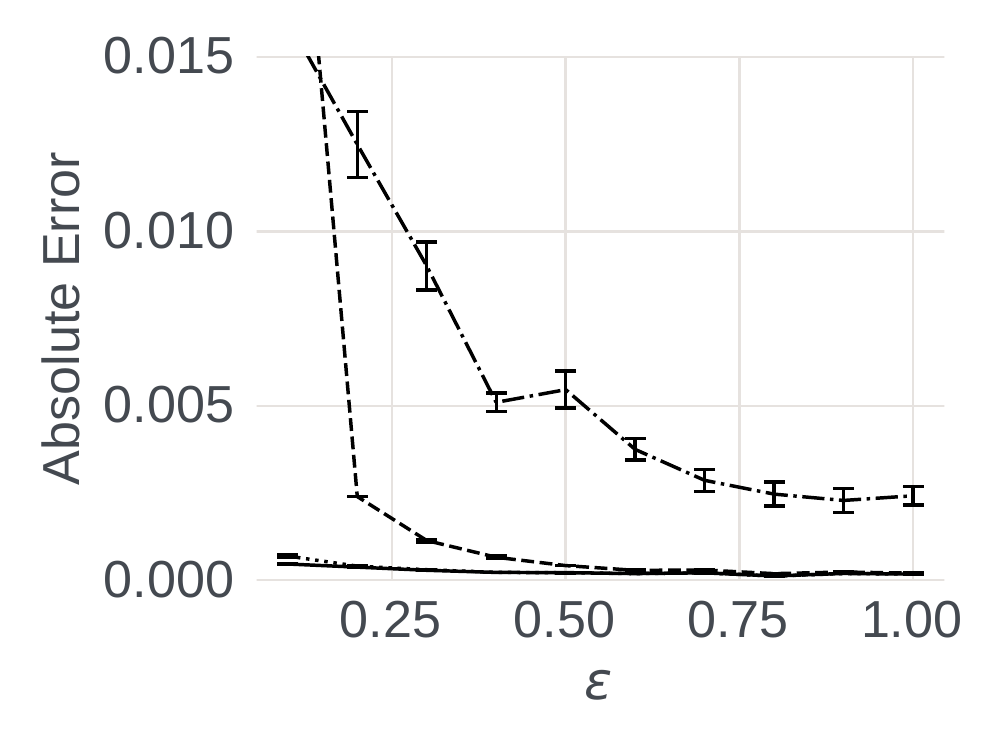}}%
\subcaptionbox{ Geometric, $B=2^6$ }{\includegraphics[width=0.33\textwidth]{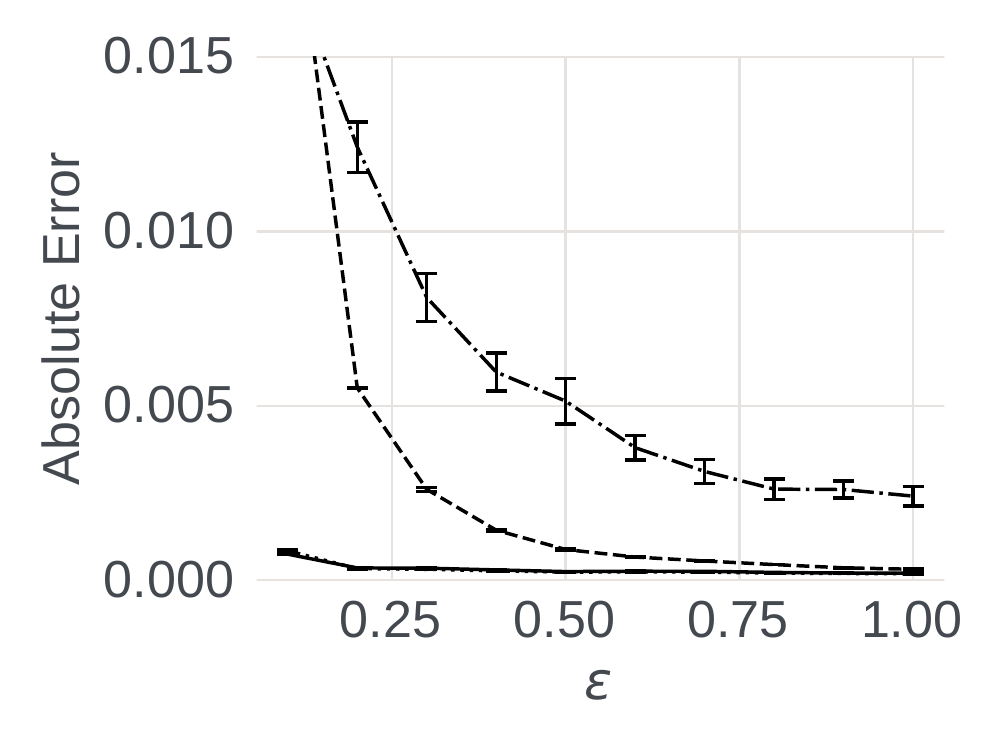}}%
\subcaptionbox{ Shakespeare, $B=2^6$ \label{fig:acc:shake:64}}{\includegraphics[width=0.33\textwidth]{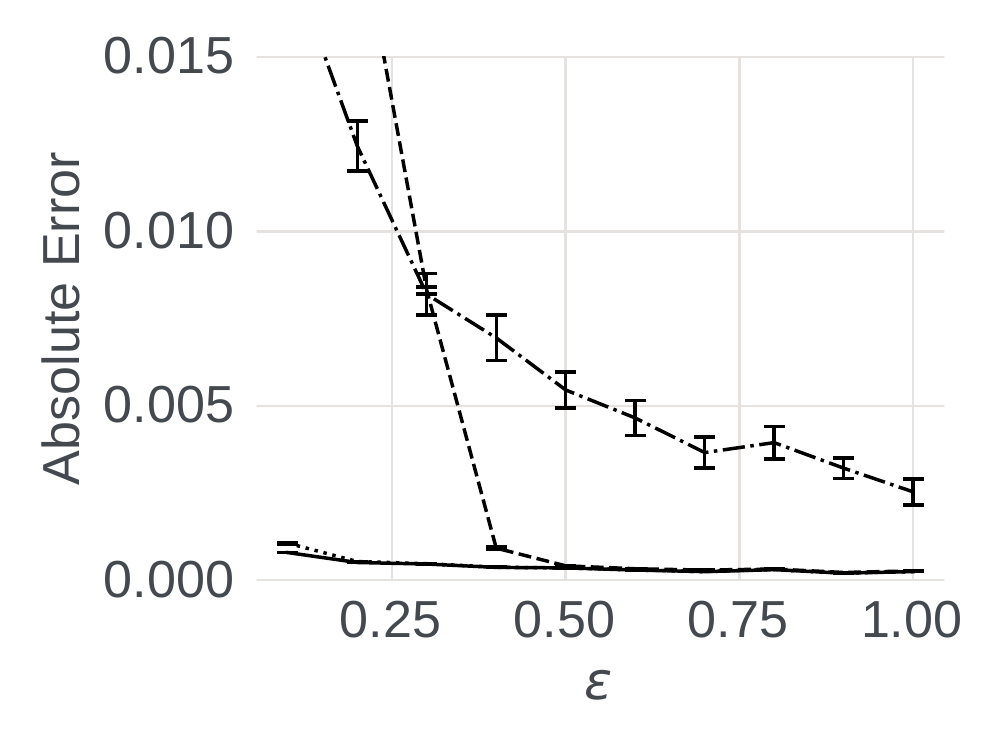}}
\subcaptionbox{ Binomial, $B=2^{10}$ }{\includegraphics[width=0.33\textwidth]{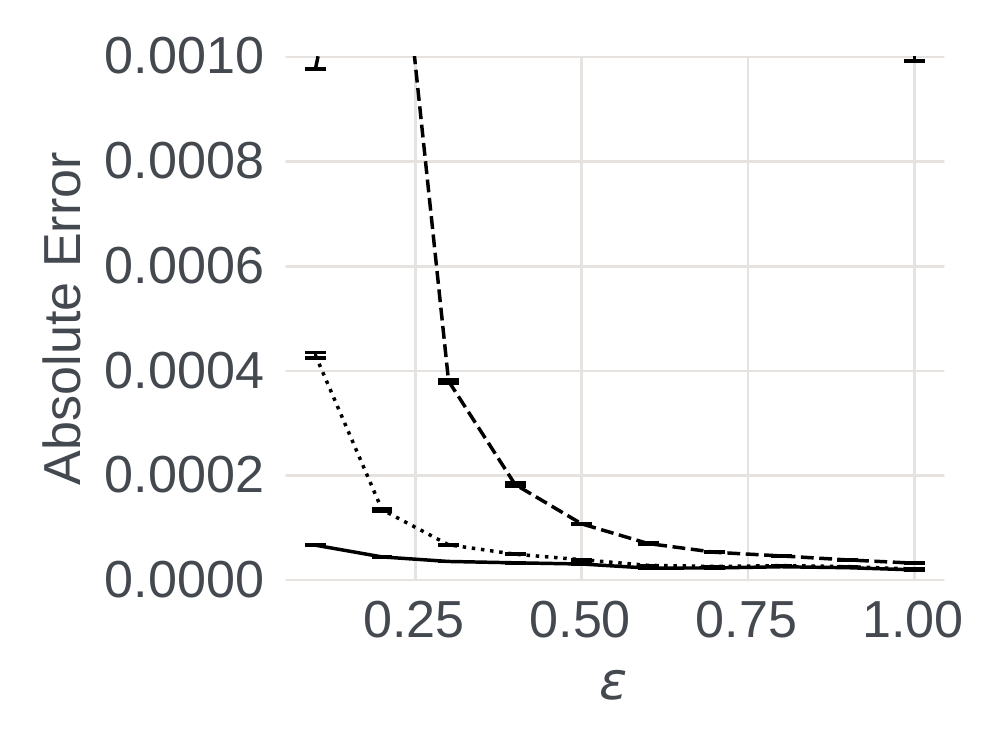}}%
\subcaptionbox{ Geometric, $B=2^{10}$ }{\includegraphics[width=0.33\textwidth]{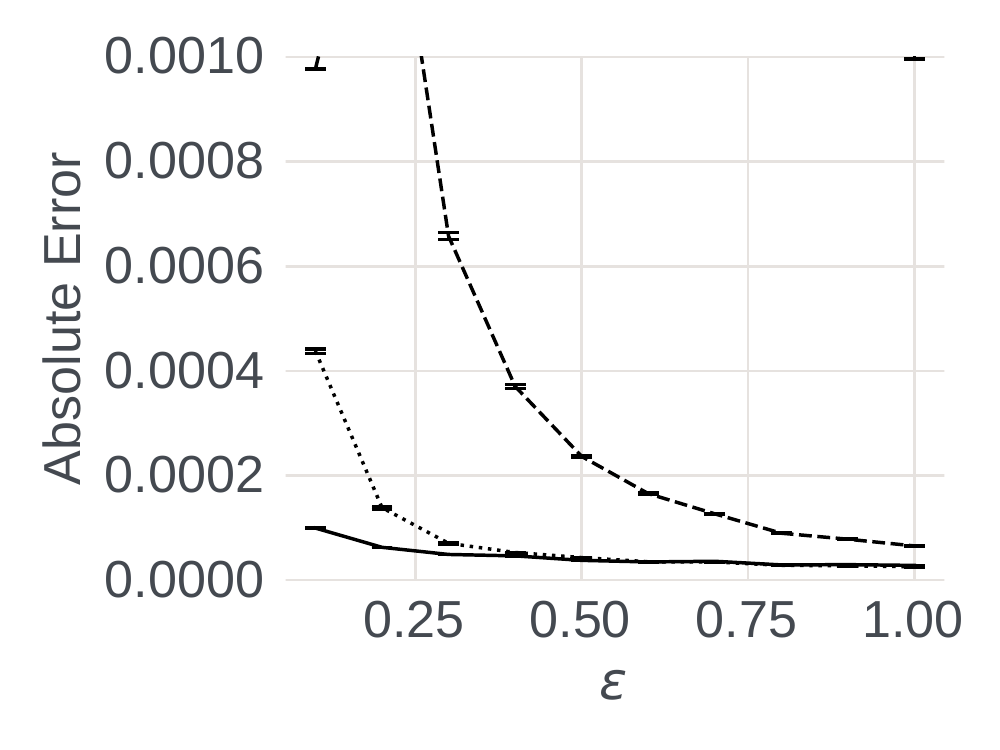}}%
\subcaptionbox{ Shakespeare, $B=2^{10}$ \label{fig:acc:shake:1024}}{\includegraphics[width=0.33\textwidth]{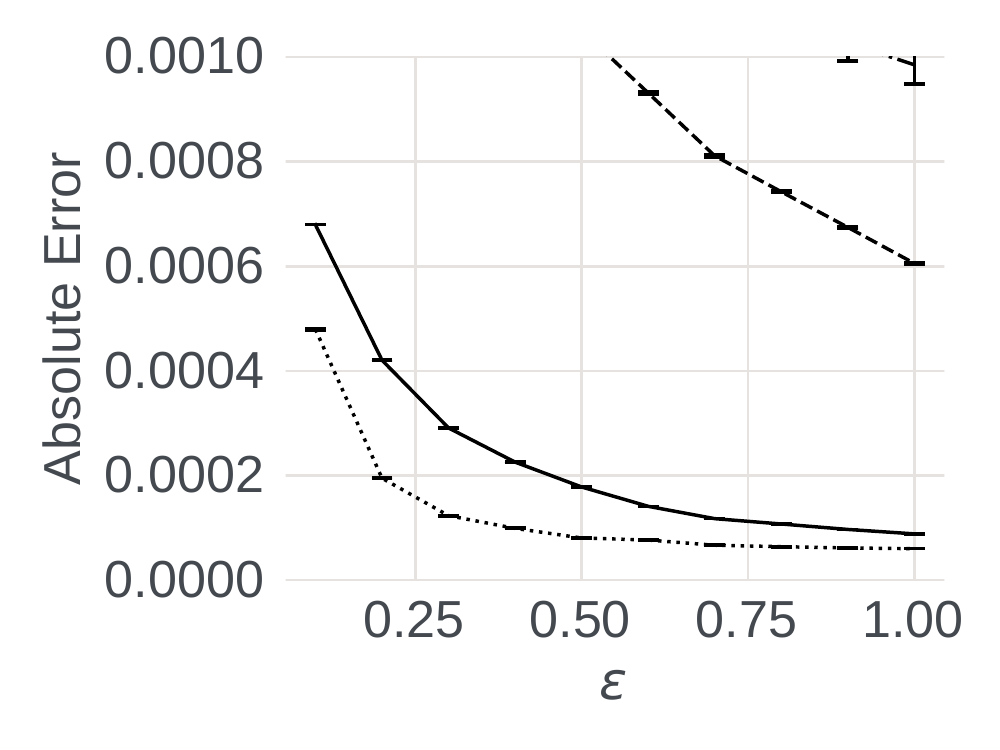}}
\subcaptionbox{ Binomial, $B=2^{14}$ }{\includegraphics[width=0.33\textwidth]{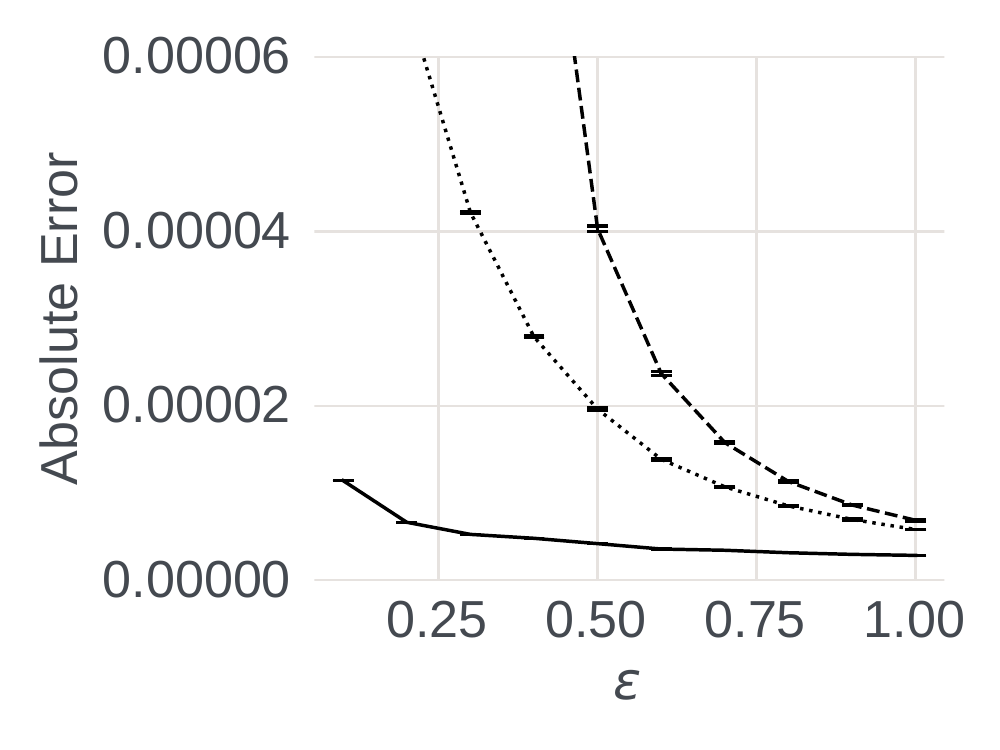}}%
\subcaptionbox{ Geometric, $B=2^{14}$ }{\includegraphics[width=0.33\textwidth]{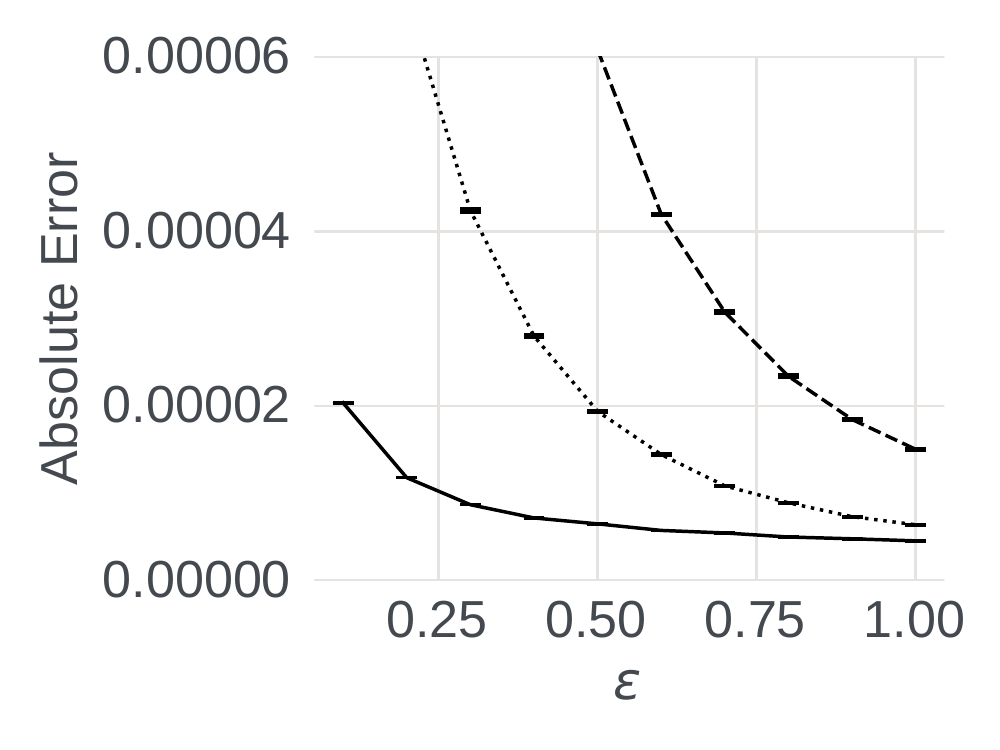}}%
\subcaptionbox{ Shakespeare, $B=2^{14}$ \label{fig:acc:shake:16384}}{\includegraphics[width=0.33\textwidth]{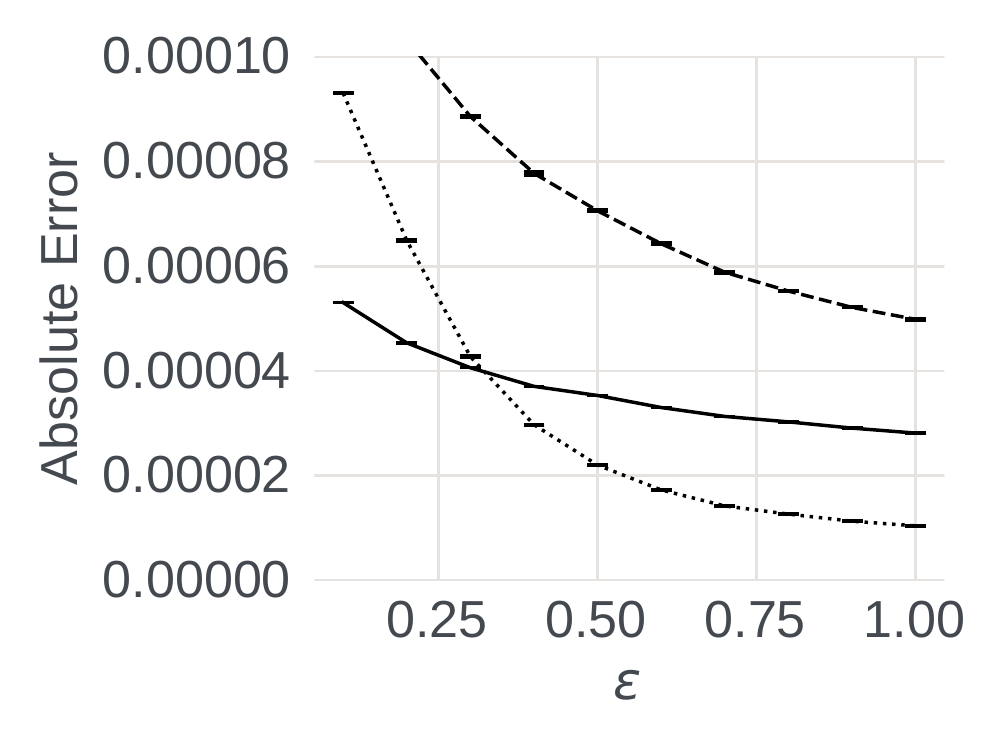}}
\caption{Accuracy results on Binomial, Geometric and Shakespeare datasets}
\label{fig:accuracy}
\end{figure*}

\begin{figure*}
\centering
\includegraphics[trim=580 25 580 25,clip,width=0.9\textwidth]{figs/hist_legend.pdf}
\subcaptionbox{ Binomial, $B=2^8$ }{\includegraphics[width=0.33\textwidth]{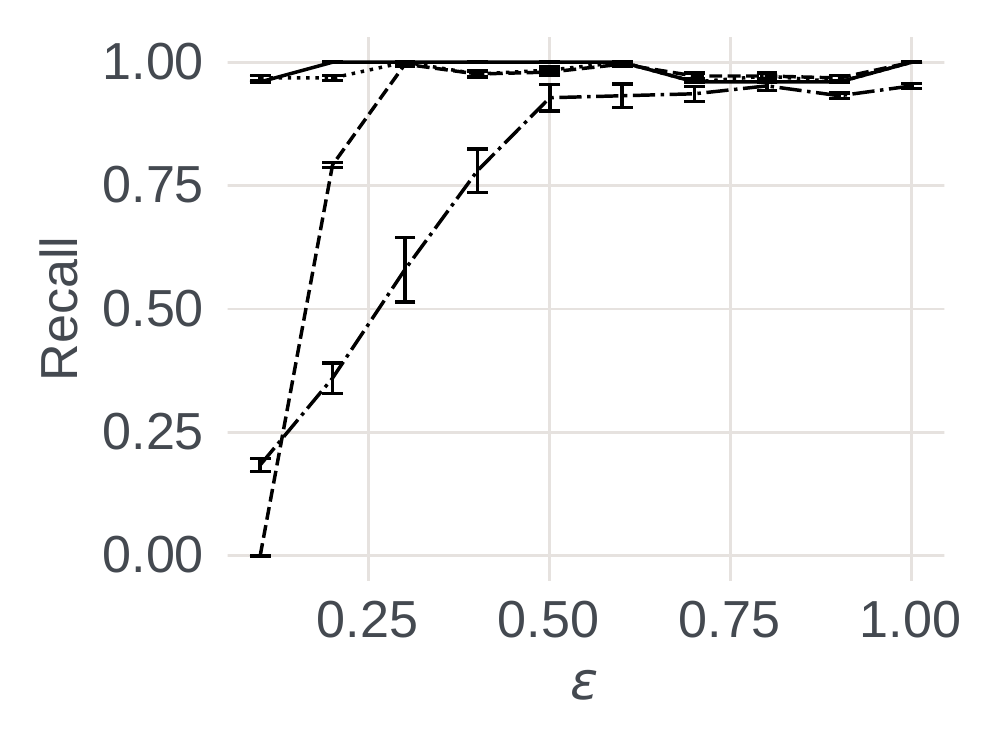}}%
\subcaptionbox{ Geometric, $B=2^8$ }{\includegraphics[width=0.33\textwidth]{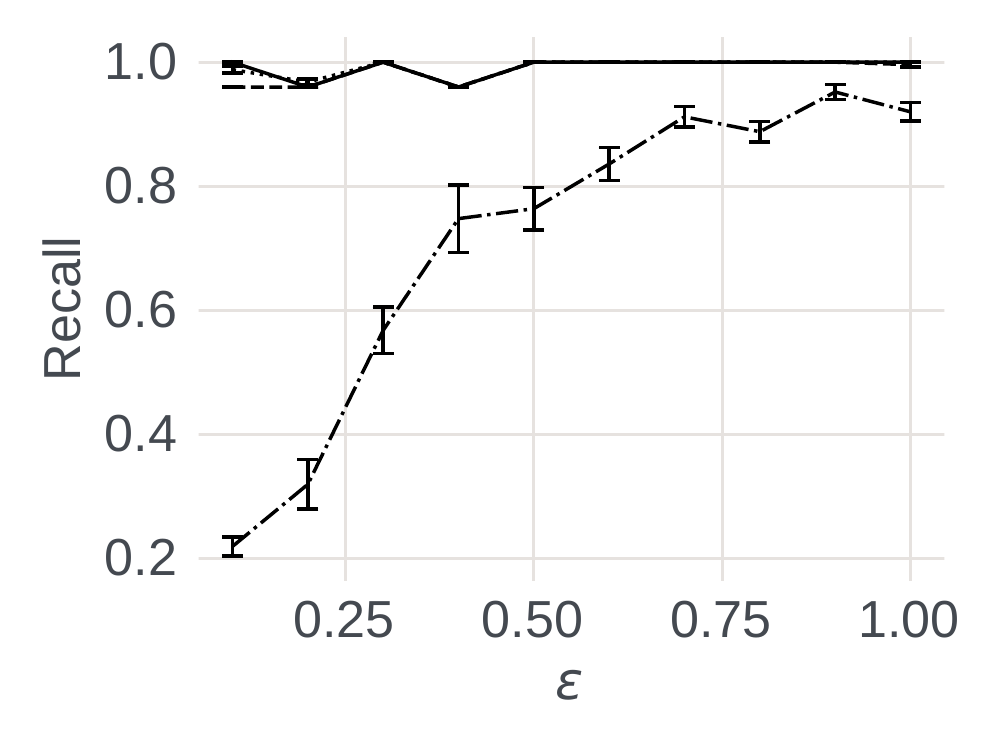}}%
\subcaptionbox{ Shakespeare, $B=2^8$ }{\includegraphics[width=0.33\textwidth]{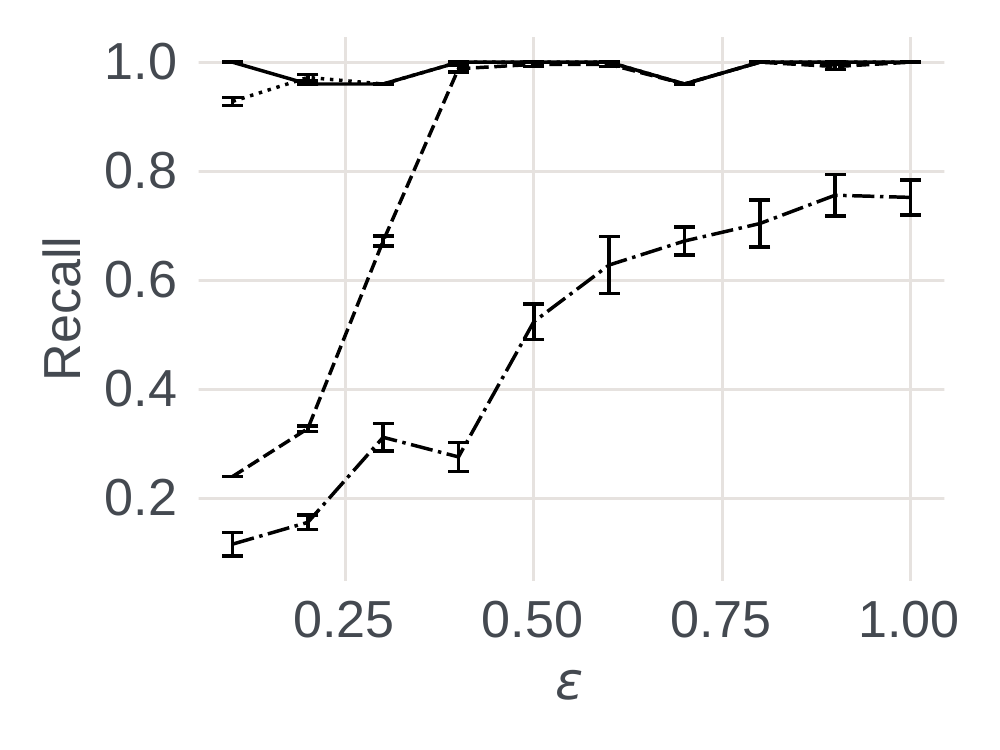}}
\subcaptionbox{ Binomial, $B=2^{12}$ }{\includegraphics[width=0.33\textwidth]{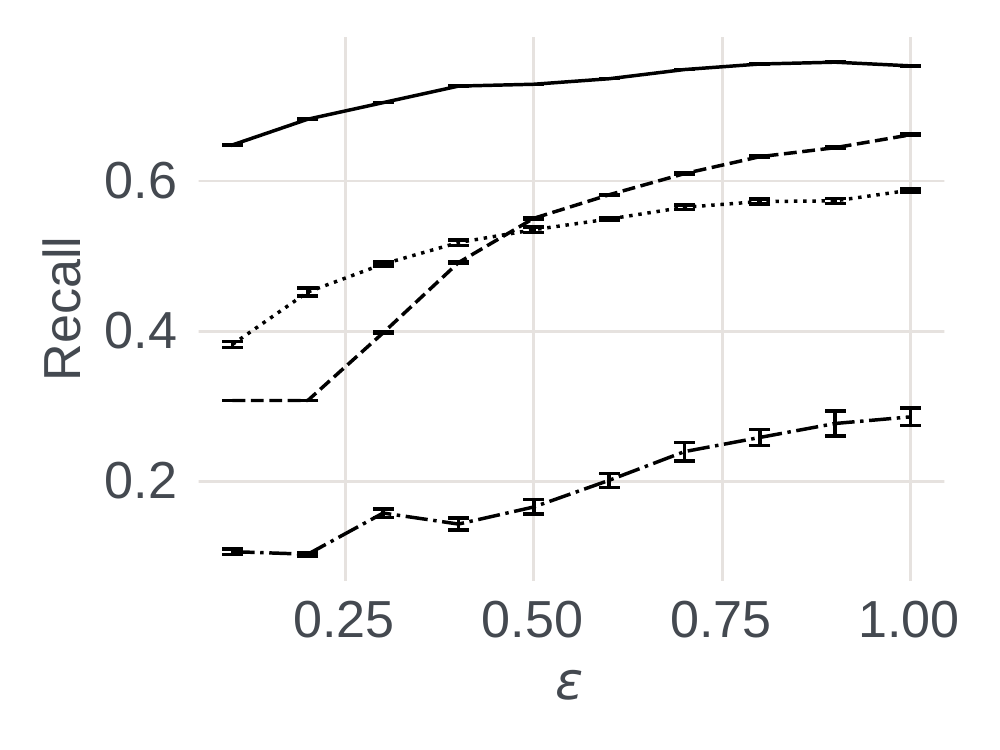}}%
\subcaptionbox{ Geometric, $B=2^{12}$ }{\includegraphics[width=0.33\textwidth]{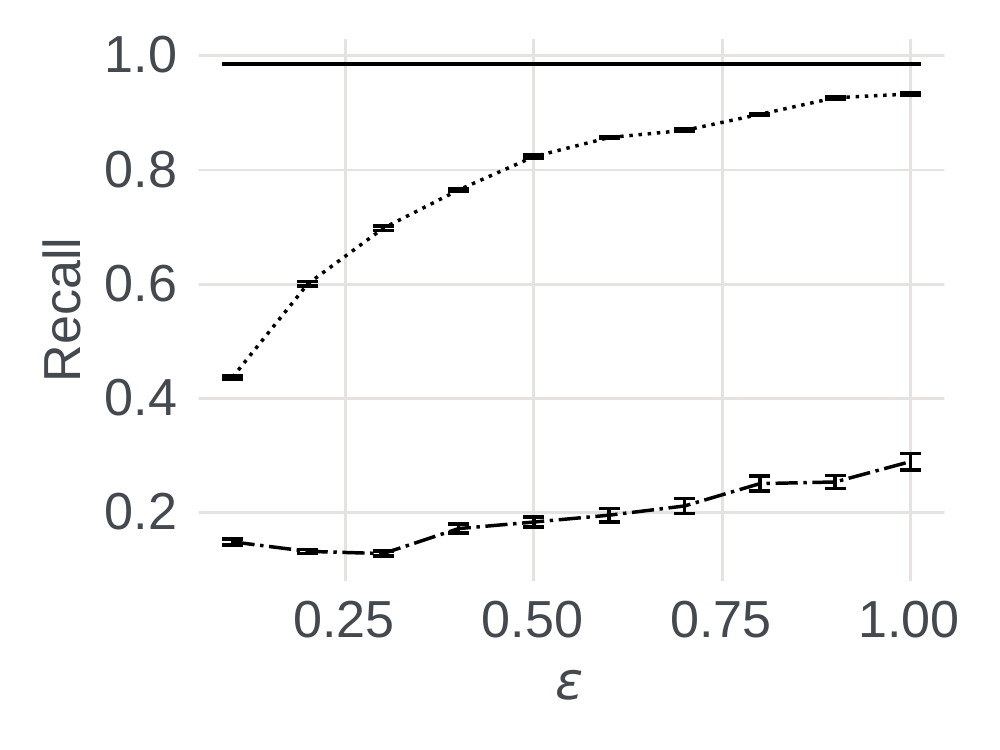}}%
\subcaptionbox{ Shakespeare, $B=2^{12}$ \label{fig:rec:shake:4096} }{\includegraphics[width=0.33\textwidth]{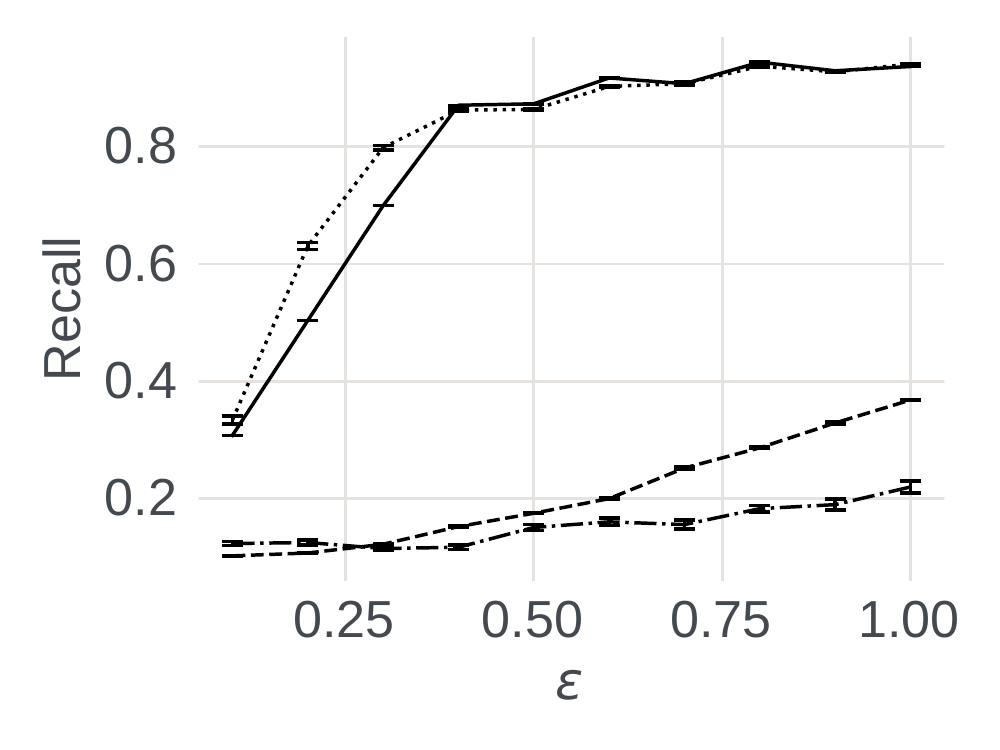}}

\caption{Top-$k$ recall results on Binomial, Geometric and Shakespeare datasets}
\label{fig:recall}
\end{figure*}

\section{\uppercase{Experiments}}
\label{sec:expts}
To validate our theoretical understanding, we performed experiments using the sample-and-threshold histogram mechanism. 
It performs sampling on a population of size $n$ for a target sample size $m$, and applies an appropriate threshold to the resulting sample, to achieve an $(\epsilon, \delta)$-DP guarantee. 
We compared against alternative mechanisms that also provide the same level of privacy when applied to the sampled set of clients: 
central differential privacy, via Laplace noise addition, local differential privacy based on Hadamard encoding of elements from the domain~\citep{hadamardresponse}, and a shuffling-approach which adds Bernoulli noise~\citep{BalcerCheu20}. 

We worked with the text from the complete works of Shakespeare\footnote{\url{http://shakespeare.mit.edu/}},
where we extract each word, consistently map the words to one of the $B$ buckets, and count the total number of words in each bucket. 
We also use synthetic data generated by distributions providing different frequency distributions: Geometric and Binomial distributions over the $B$ cells of the histogram. 
For the Binomial data, each client draws from the Binomial distribution with $n=B$ and $p=0.5$ to choose a histogram bucket. 
For the Geometric data, each client draws from a Geometric distribution with $p=1/\sqrt{B}$ to pick a histogram bucket. 
These parameters are chosen to model the non-uniform frequency distributions seen in practice, where the most popular items occur approximately 1-5\% of the time. 

We experimented with a range of privacy parameters $\epsilon$, $\delta$, histogram sizes $B$, and population sizes $n$. 
We pick a default $\alpha = 1/6$ and $\delta = 10^{-8}$, which yields a threshold $\threshold = 20$. 
We simulate a population of $n = 10^6$ clients, 
and measure the accuracy of recovering the frequencies for each mechanism. 
We compare the absolute difference of the estimated frequencies to those from the full population, 
and also measure the recall for the top-$k$ heaviest buckets for $k=B/10$, i.e., the largest 10\% of frequencies. 
In the plots, we focus on showing results for the range of $\epsilon = 0.1$ (high privacy) to $\epsilon = 1.0$ (medium privacy) regimes, consistent with the range where all the mechanisms have privacy guarantees.
We vary the size of the histograms ($B$) from tens up to tens of thousands. 
Error bars show the standard error over 10 repetitions of each mechanism. 
Plots for other parameter settings are withheld for brevity, but support the same conclusions.  

\paragraph{Accuracy results.}
Our results on accuracy are shown in Figure~\ref{fig:accuracy}. 
Each row shows results for a different histogram size, from small ($B=2^6$), to large ($B=2^{14}$); 
each column shows results on a different dataset (Binomial, Geometric or Shakespeare data). 
The y-axis shows absolute error, expressed as a fraction of the total input size.
We want this to be as low as possible, and ideally much smaller than 0.1\%, say. 

Some results immediately stand out: the results from local differential privacy are much weaker, and frequently the error is sufficiently large that the line does not appear on the plots (similar results were seen for other choices of frequency oracle, such as direct encoding and unary encoding~\citep{wangetal}---we use the Hadamard encoding as it obtained the best accuracy for these experiments).  
This is consistent with our understanding of LDP, and further motivates the desire to achieve accuracy closer to the centralized case in federated settings. 
The approach from the shuffle model, where each client adds Bernoulli noise to each cell of the histogram (i.e., for each cell they report a 1 value with some probability $q$) incurs higher error for small $\epsilon$ (where more noise is added by the sampled clients). 
The gap is larger as the size of the histogram increases, since there are more chances for cells to incur more noise.  
Most intriguingly, the approach of adding Laplace noise, which is the gold standard in the centralized case, does not obtain the least error in this setting.  
Rather, the sample and threshold approach, which does not add explicit noise, but just removes small sampled counts, often achieves less error, particularly for small $\epsilon$, where the magnitude of the Laplace noise is larger. 
This is more pronounced for larger histograms. 
The exception is for the Shakespeare data for larger histograms (Figures~\ref{fig:acc:shake:1024} and~\ref{fig:acc:shake:16384}). 
Here, the combination of skewed data, and smaller sample sizes for smaller $\epsilon$, means that only a fraction of histogram buckets pass the threshold (often, fewer than 10\% of buckets). 
These buckets contribute little to the distribution, but
while the sample-and-threshold improves over shuffling, it does not reach the accuracy of central noise addition when there are many infrequent items. 

Last, we note that the magnitude of the error decreases as the histogram size increases. 
This is in part since the magnitude of the bucket frequencies decreases, and we are showing the (mean) error per bucket. 
As a sanity test, we also computed accuracy of the trivial approach of reporting zero for each bucket. 
The error for this approach falls above the range of each graph plotted, giving reassurance that we are achieving non-trivial accuracy for the histogram problem. 

\paragraph{Recall results.}
To better understand the ability of the different approaches to capture the high counts (as needed for finding heavy hitters), we measure the recall of the top-$k$ items, for $k = B/10$. 
That is, we test whether the histogram correctly reports the (true) top-10\% of items among the 10\% of largest items recovered. 
Figure~\ref{fig:recall} shows the results across different datasets. 
For moderate sized histograms ($B=2^{8}$), the sample and threshold approach achieves close to perfect recall for all datasets.  
Other methods are comparable, but weaker for small $\epsilon$.  
Again, it is the Shakespeare data for a larger histogram that presents the greatest challenge (in Figure~\ref{fig:rec:shake:4096}).  
Here, the same issue as above affects sample and threshold: 
a large fraction of small frequencies mean that these do not meet the threshold for the sample size. 
Accepting a larger $\epsilon$, or working over a larger population to obtain a larger sample size would be needed to improve the recall. 
However, it could be argued that items missed are not very significant: already at $\epsilon = 0.2$, the threshold applied means that only items with frequency less than 0.1\% are likely to be dropped. 
\section{\uppercase{Related work}}
\paragraph{Histograms.}
Due to their broad applications, histograms are one of the most heavily studied tasks in differential privacy (DP). 
One of the first DP results is that a private histogram can be created by adding independent Laplace noise to each entry of the exact histogram~\citep{Dwork06,DworkR14}.
For large domains, an $(\epsilon, \delta)$-DP guarantee can be obtained by applying a threshold to the noisy counts, and omitting any histogram entries whose (true) count is zero, which preserves the sparsity of the input~\citep{Korolova:Kenthapadi:Mishra:Ntoulas:09,Bun:Nissim:Stemmer:18,Balcer:Vadhan:18}. 
For multi-dimensional data, histograms of low-degree marginal distributions can be created via noise addition to the Hadamard transform of the data~\citep{BarakCDKMT07}. 
These results assume a given set of histogram bucket boundaries; 
\cite{XuZXYY12}
considered choosing bucket boundaries privately to minimize squared error. 
The histogram problem has also been heavily studied in the local model of DP, where each individual adds noise to their input independently. 
Here, histograms are often implemented via `frequency oracles', and used to identify frequent items from the input~\citep{bassilysmith}. 
Optimized constructions make use of hashing~\citep{wangetal} and Hadamard transforms~\citep{hadamardresponse} to minimize the variance of the estimate. 
More recently, results are shown in the shuffle model, where messages from individuals are anonymized by a ``shuffler'', so the analyst sees only the multiset of messages received without attribution~\citep{ESA20}. 
For shuffling with a fixed privacy level $\epsilon$, accuracy bounds closer to the central case are achievable by introducing  small amounts of random noise from each participant~\citep{BalcerCheu20,DUMP20,Ghazi21}. 

\paragraph{Heavy hitters.}
The problem of finding the most frequent items from a collection is a core analytics task that supports a range of objectives, from simple popularity charts, to instantiating complex language models.  
Due to the sensitivity of data used within these applications, it is  necessary to apply strong privacy protections to the data. 
There have been multiple efforts to address this problem in the Local DP setting~\citep{bassilysmith,wangetal,wangConsistency, rappor, appledp, practicalhh} and shuffle model~\citep{Ghazi21}.
The closest work to ours is recent work on Federated Heavy Hitters discovery~\citep{TrieHH},
which describes an $(\epsilon, \delta)$-DP algorithm to collect information from a set of distributed clients, who each hold a (private) item.  
We can treat these items as strings of characters over a fixed alphabet. 
The algorithm proceeds in a series of $L$ rounds to build up a trie describing the frequent items among the client population.  
In each round, the server contacts a random sample of $m$ clients, and shares the current trie with them. 
Each client replies if its item extends the trie, and if so the client ``votes'' for the prefix that its item extends, along with the next character. 
The server receives these votes, and tallies them.  
Popular prefixes are added to the trie, and are candidates for further extension in the next round. 
The procedure stops after the trie has been built out to $L$ levels, or if the trie cannot be extended beyond a certain level. 

\paragraph{Quantiles and range queries.}
The quantiles of a distribution give a compact description of its (one-dimensional) CDF, generalizing the median. 
The problem has also been studied in the central, local and shuffled models. 
Many solutions first solve range queries, then reduce quantile queries to  range queries. 
\cite{XiaoWG10} propose using the Haar wavelet transform with noise, 
while \cite{QardajiYL13} use hierarchical histograms. 
\cite{CormodeKS19} compare both methods in the local setting and observe similar levels of accuracy. 
In the shuffle model, quantiles are addressed via frequency histograms in the work of \cite{Ghazi21}. 

\paragraph{Sampling and DP.}
It is well-known that sampling can be used to amplify the guarantees of differential privacy when combined with a DP mechanism on the sample: 
\cite{BalleBG20} show results for Poisson sampling, and fixed-size sampling with and without replacement,
while \cite{Imola21} study privacy amplification when sampling according to differentially private parameters. 
By contrast, we consider mechanisms where sampling in isolation (with a threshold) provides the DP guarantee directly. 
This idea is inspired by~\cite{TrieHH}, which materializes a set of items based on sampling and thresholding. 
The key advance in our work is to show that we can output the sampled frequencies as well as the sampled items, and hence produce private histograms. 
Also similar to our work is that of 
\cite{Li:Qardaji:Su:12}, who combine sampling with $k$-anonymization to achieve a DP guarantee.  
Here, we are able to give tight bounds to guide how to set the threshold and sampling rate based on target $\epsilon$ and $\delta$ values. 
Our work complements other efforts in the federated setting to achieve privacy guarantees with a restricted set of operations---for instance, \cite{KairouzM00TX21} seek to perform federated learning via noise addition \textit{without} sampling. 


\section{\uppercase{Concluding Remarks}}

We have shown how the sample-and-threshold approach can be applied to the fundamental problem of private histogram computation, and related tasks like heavy hitter and quantile estimation. 
The key technical insight is that sampling a large enough number of indistinguishable examples introduces sufficient uncertainty to meet the differential privacy guarantee. 
The uncertainty introduced due to sampling can be viewed as a forced randomized response (where 0 values are forced to be reported as zeroes) or ``negative noise'', in contrast to other mechanisms that add explicit noise.  
It will be interesting to try to understand the relations between these different forms of noise.

As with other works on private histograms, we assume that the bucket boundaries of the histogram are given. 
Adaptive division of the histogram buckets is possible, as seen in the TrieHH++ protocol. 
Nevertheless, this approach can give poor results in extreme cases, such as when the bulk of the data resides in a very small fraction of the input domain. 

It is natural to consider what other computations might benefit from this sample-and-threshold approach. 
Direct application of the technique makes sense when many users hold copies of the same value. 
Hence, it is not well-suited to questions like finding sums and means of general distributions, unless we additionally apply some rounding and noise addition to input values first. 
The approach may be of value for more complex computations, such as clustering or outlier removal, where dropping rare items is a benefit, or tasks where we seek to discover descriptions of patterns in the data that have large support, such as frequent itemsets. 

\paragraph{Acknowledgements.}
We thank
Ari Biswas, Dzimtry Huba, Igor Markov, Ilya Mironov, Harish Srinivas, and Kaikai Wang
for useful comments and feedback.